\documentclass[10pt]{article}

\usepackage{graphicx}
\usepackage{dcolumn}
\usepackage{bm}

\begin{document}
\begin{center}
{{ \bf Weibull model of Multiplicity Distribution in hadron-hadron collisions }} \\
\bigskip
{ Sadhana Dash$^*$\footnote{email: sadhana@phy.iitb.ac.in}  and  Basanta~K.~Nandi} \\
{$^*$Indian Institute of Technology Bombay, Mumbai, India } 
\bigskip
\end{center}

\begin{abstract}
We introduce the Weibull distribution as a simple parametrization of
charged particle multiplicities in hadron-hadron collisions at all
available energies, ranging from ISR energies to the most recent LHC
energies. In statistics, the Weibull distribution has wide applicability
in natural processes involving fragmentation processes. This gives a
natural connection to the available state-of-the-art models for
multi-particle production in hadron hadron collisions involving QCD
parton fragmentation and hadronization. 
\end{abstract}


\section{Introduction}

Inclusive charged particle multiplicity distribution had been
extensively  studied  for a long time in hadronic collisions. The
multiplicity distribution characterizes the hadronic final state and
provides important insights on the production mechanism of muti-particle final states. 
In particular, they are sensitive to the underlying QCD dynamics in
hadron-hadron collisions. 
The previous phenomenological studies of multiplicity distributions at
various  energies have been done in terms of parameters of  the Negative
Binomial Distribution (NBD) function. Although the interpretation of the NBD in
terms of particle production is not fully understood, it
remarkably described  the data at lower energies \cite{1}. However,
deviations from the NBD were observed at higher energies \cite{1,2,3}
which puts constraints on its universal applicability for wider range of
energies.  It was followed by two-component model of particle
production, interpreting one as the soft and the other as the
semi-hard component \cite{4}.  This led to the description of the data
by the weighted combination of two NBD's. The multiplicity distributions  at
LHC energies and for all pseudorapidity intervals were well
described by this approach \cite{2,5}. However, it was pointed out that
the excellent description of  multiplicity distributions of hard QCD
events in larger $\eta$ window of $p$-$p$ collisions at 7 TeV
contradicts the very  concept of two component model \cite{5}.
A very detailed discussion on multiplicity measurements and various
approaches can be found at \cite{6}.
In this work, we introduce a statistical distribution, namely, the
Weibull distribution \cite{7} which we found to describe the $p$-$p$($p$-$\overline{p}$) data  
at all available energies ranging from ISR to LHC. In addition, this
distribution  gives an appealing physical interpretation.

\section{Weibull distribution }
Many evolving systems in nature exhibit skewed distributions and 
among them Weibull and lognormal distributions appear more
ubiquitously in a variety of systems \cite{7,8}. In particular, the Weibull
distribution is widely used to describe size distribution obtained in
diverse fields such as material fragmentation \cite{9} , cloud droplets
\cite{10}, biological systems \cite{11,12}, wind speeds \cite{13} and
extreme value statistics \cite{14,15}  etc.
Since final state particle  multiplicity can be seen as evolved from
initial hadron-hadron collisions,  one can  expect to use the  Weibull
distribution to describe the probability of producing charged particles.

The probability distribution of a continuous random variable `$n$' in terms of two
parameter Weibull distribution 
is given by 
\begin{equation}
P \left( n , \lambda , k\right) = \frac{k}{\lambda} \left(
  \frac{n}{\lambda} \right)^{k-1}  e^{-{(n/\lambda)}^{k}}
\end{equation}  

where $k$ is the shape parameter and $\lambda$ is the scale parameter of 
the distribution. 
The mean of the distribution is given in terms of $\lambda$ and k as, \\
\begin{equation}
  \langle n \rangle  =   \lambda \Gamma(1 + \frac{1}{k})  
\end{equation}  

The description of a physically based derivation of  Weibull 
distribution with respect to fragmentation processes can be found at 
\cite{7}.  In this particular work \cite{7}, it was shown that the result of a single event 
fragmentation leading to a branching tree of cracks in the material
that show fractal behaviour can be described by Weibull distribution. This can be 
related to the mass distribution  or particle number distribution developed by 
fragmentation process.  It was also shown that a particular mass 
distribution closely resembles the lognormal distribution which was 
quite popular for sometime describing fragmentation distributions
\cite{7}. The lognormal distribution also described well the particle 
multiplicities at ISR energies \cite{16}.

The underlying mechanism of particle production in hadronic collisions
as given by current models of particle production is based on the
fragmentation of partons into observed hadrons. Irrespective of
particular details, most of the models of multi-particle production
contain an evolution composed of various steps that are based on
previous intermediate steps and are influenced by the same. The steps
involving hard and semi-hard processes are well explained by
perturbative quantum chromodynamics (QCD)  \cite{17,18,19} while
models for soft interactions involve large class of string
fragmentation models \cite {20, 21}.   Regardless of differences in
details, the cascade nature of models involving fragmentation of
partons is apparent and thus one can use Weibull distribution to model 
the basic multiplicity  distribution in hadron-hadron collisions.
This also corroborates to the clan model involving particle
cascades as mechanism of particle production \cite{2,4}.
\section{Multiplicity Distribution and Weibull Parameters}

In the present scenario where we have experimental data on
multiplicity distributions in widest range of energies and
pseudorapidity intervals, it is worth trying to check whether we can 
parametrize the data in terms of Weibull parameters.   
We identify the variable $n$  with the charged particle multiplicity and
perform the fits to the data points  using the chi-square minimization
method. 
We fitted the multiplicity distribution in $p$-$\overline{p}$ and $p$-$p$ collisions as
measured by UA5 experiment at SPS energies \cite{1} and by  CMS
experiments \cite{17} at LHC energies, respectively. 
Figure 1(a) shows the Weibull fits to the multiplicity distribution at
mid-rapidity for the $p$-$\overline{p}$ collisions as measured by  UA5
experiments at 200 GeV, 540 GeV and 900 GeV. The fits to the recent  
multiplicity distributions in $p$-$p$ collisions as measured by the CMS
experiment at  900 GeV, 2.36 TeV and 7 TeV  are shown in Figure
1(b). As can be seen from the figures, the Weibull fit gives an excellent
description of the data at all energies. The  values of the parameters and the
chi-square of the fit is tabulated in Table 1. We can easily compare  that the
parameters obtained from the fits at $p$-$\overline{p}$  and $p$-$p$
collisions at 900 GeV  (from two different experiments) are comparable  within errors.
\begin{figure}
  \includegraphics[scale=0.31]{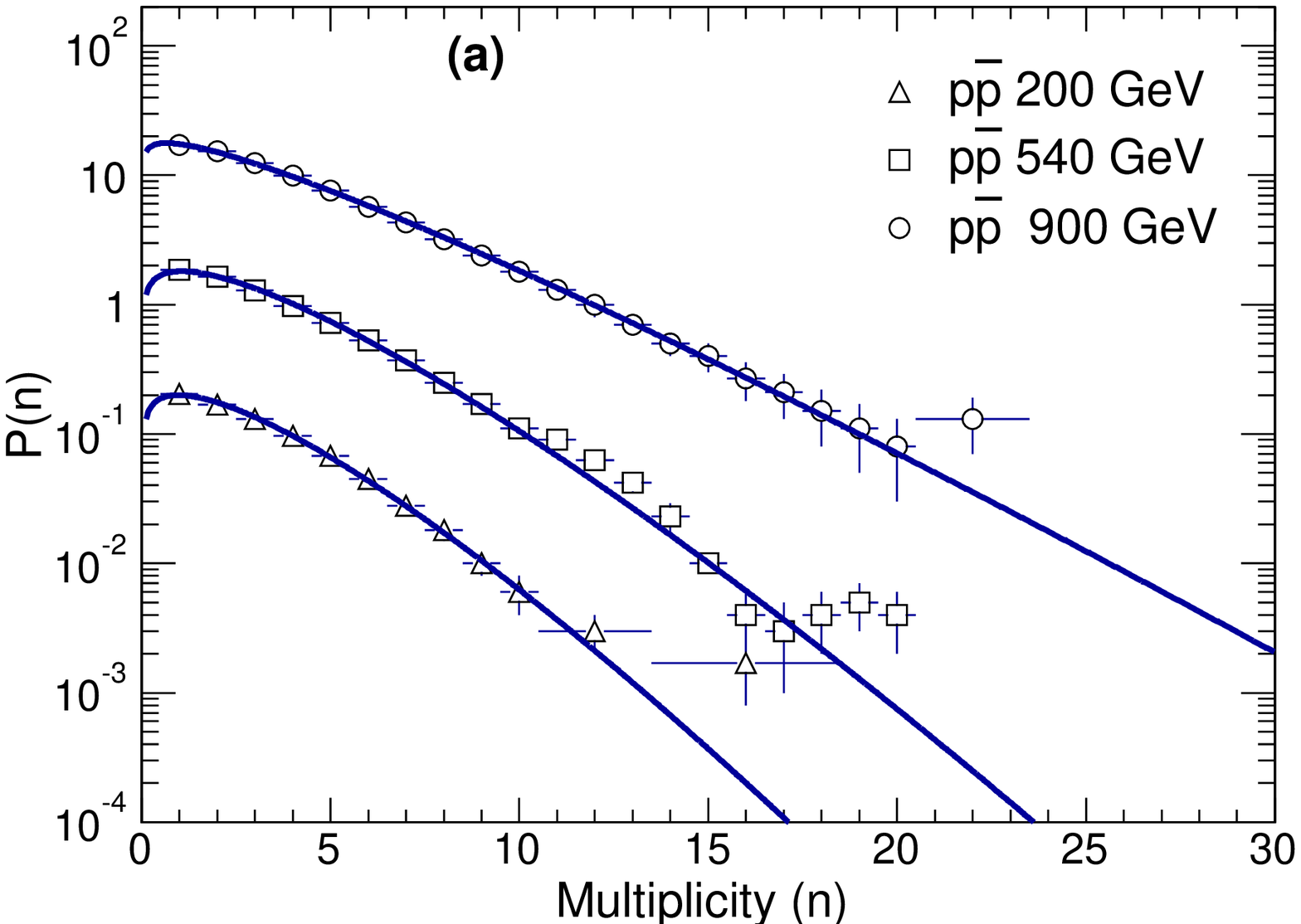}
  \includegraphics[scale=0.31]{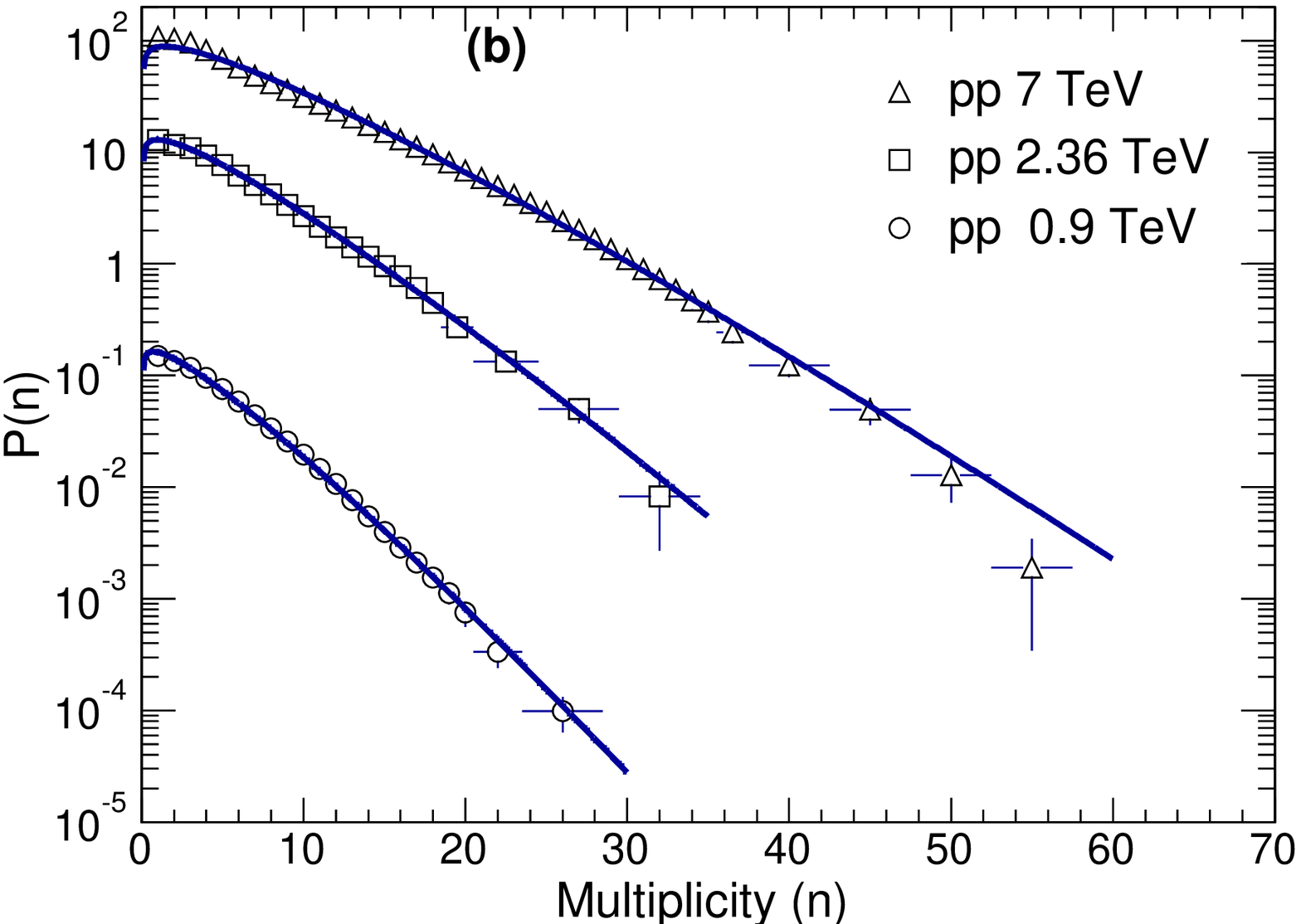}
  \caption{ (a) UA5 measurement of charged particle multiplicity
    distribution in $p$-$\overline{p}$ collisions at mid-rapidity at
    $\sqrt{s_{NN}}$ = 200 GeV, 540 GeV and 900 GeV \cite{1}.(b) CMS measurement
    of multiplicity in $p$-$p$ collisions at mid-rapidity at  $\sqrt{s_{NN}}$
    = 0.9 TeV,  2.36 TeV and 7 TeV \cite{22}.  The solid line represents the
    Weibull fit to the data points. The data points for a given energy 
    are appropriately scaled for better visibility. }
\label{midrap}
\end{figure}
Table 1 lists all the  parameters obtained from the Weibull fit
together with the chi-square and extracted mean multiplicity for different collision systems studied.

\begin{table*}[!h]
\centering
\begin{tabular}{|c|c|c|c|c|}
\hline
\hline
 Collision systems  &  $k$  & $\lambda$  & $\chi^{2}/ndf$ & $\langle n \rangle$\\
\hline
$p$-$\overline{p}$~(200 GeV)   & 1.27 $\pm$ 0.04 &  3.17  $\pm$0.07 &
0.54  & 2.93 $\pm$ 0.029\\
\hline                 
$p$-$\overline{p}$~(540 GeV)   & 1.26 $\pm$ 0.01 &  3.58  $\pm$ 0.04   &
1.66 & 3.32 $\pm$ 0.015\\
\hline
$p$-$\overline{p}$~(900 GeV)    & 1.11  $\pm$ 0.01 &  4.07  $\pm$ 0.08
& 0.33 & 3.91 $\pm$ 0.015 \\
\hline
$p$-$p$~(0.9 TeV)    & 1.13 $\pm$ 0.01 &  4.17 $\pm$  0.06 & 0.03
&3.98 $\pm$ 0.018\\
\hline
$p$-$p$~(2.36 TeV)    & 1.14  $\pm$ 0.03&  5.41  $\pm$ 0.15 & 0.22&
5.15 $\pm$ 0.036\\
\hline
$p$-$p$~(7 TeV)    & 1.15  $\pm$ 0.01 &  7.35 $\pm$ 0.16 & 0.85 & 6.98
$\pm$ 0.026\\

\hline
\hline
\end{tabular}
\caption{ List of parameters and chi-square values of the Weibull fit 
  to the data for various collision systems.} 
\end{table*}
At LHC energies, the multiplicity distribution as measured by  LHCb
experiment covers the widest range of  pseudorapidity intervals at 7
TeV \cite{23}. The single NBD distribution could not give a good description of the
data in different pseudorapidity intervals \cite{5}. 
Figure 2(a) shows the fit for the LHCb data for various pseudorapidity
intervals in $p$-$p$ collisions at 7 TeV. 
The data are remarkably described by the Weibull distribution in all the 
pseudorapidity intervals.
The applicability was also checked at lower energies measured by  ISR
experiment \cite{24} as shown in Figure 2(b). This figure provides an excellent
comparison where we observe that the Weibull distribution successfully
explains the data at two diverse energy domains.
\begin{figure}
\includegraphics[scale=0.31]{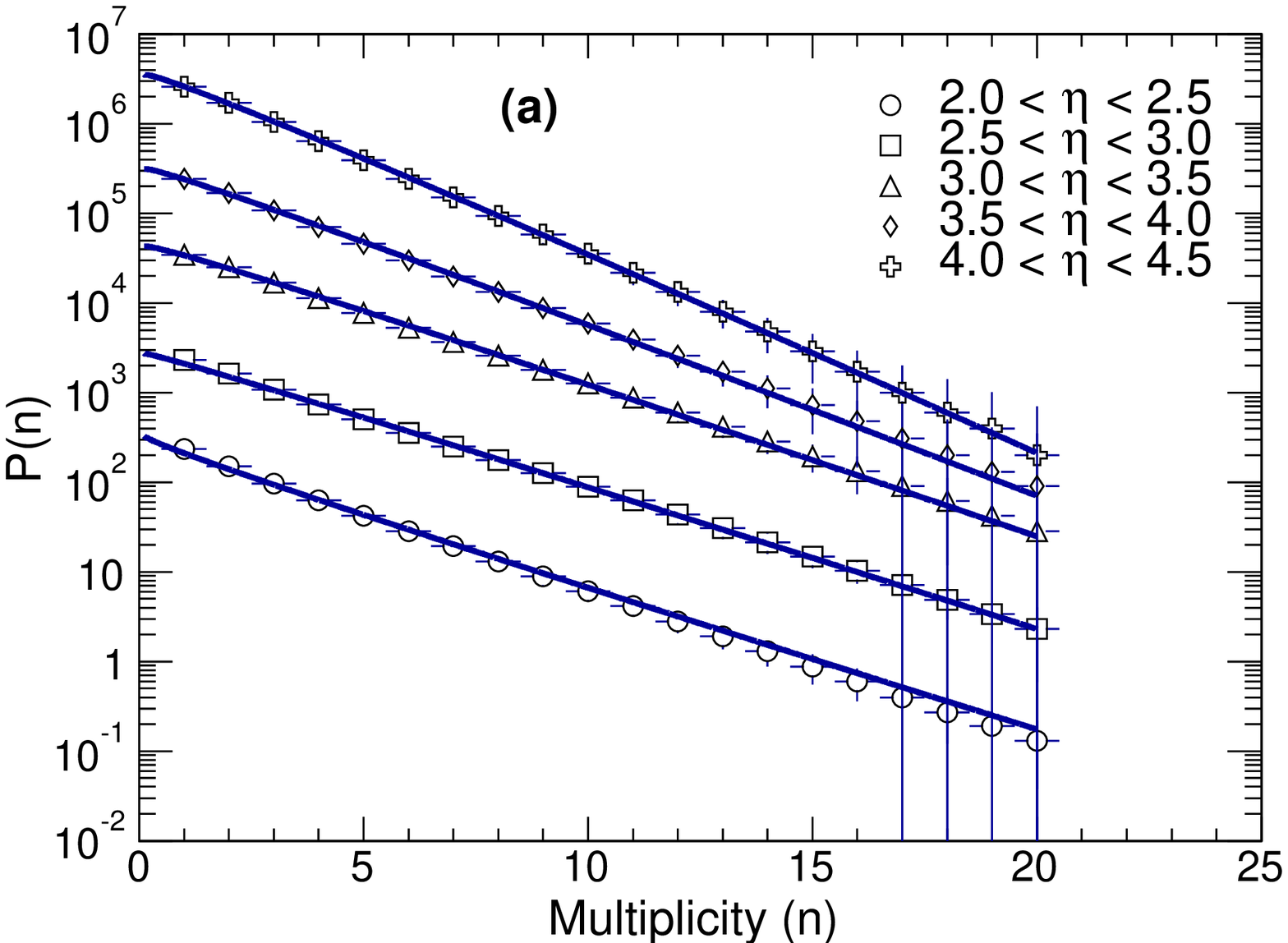}
\includegraphics[scale=0.31]{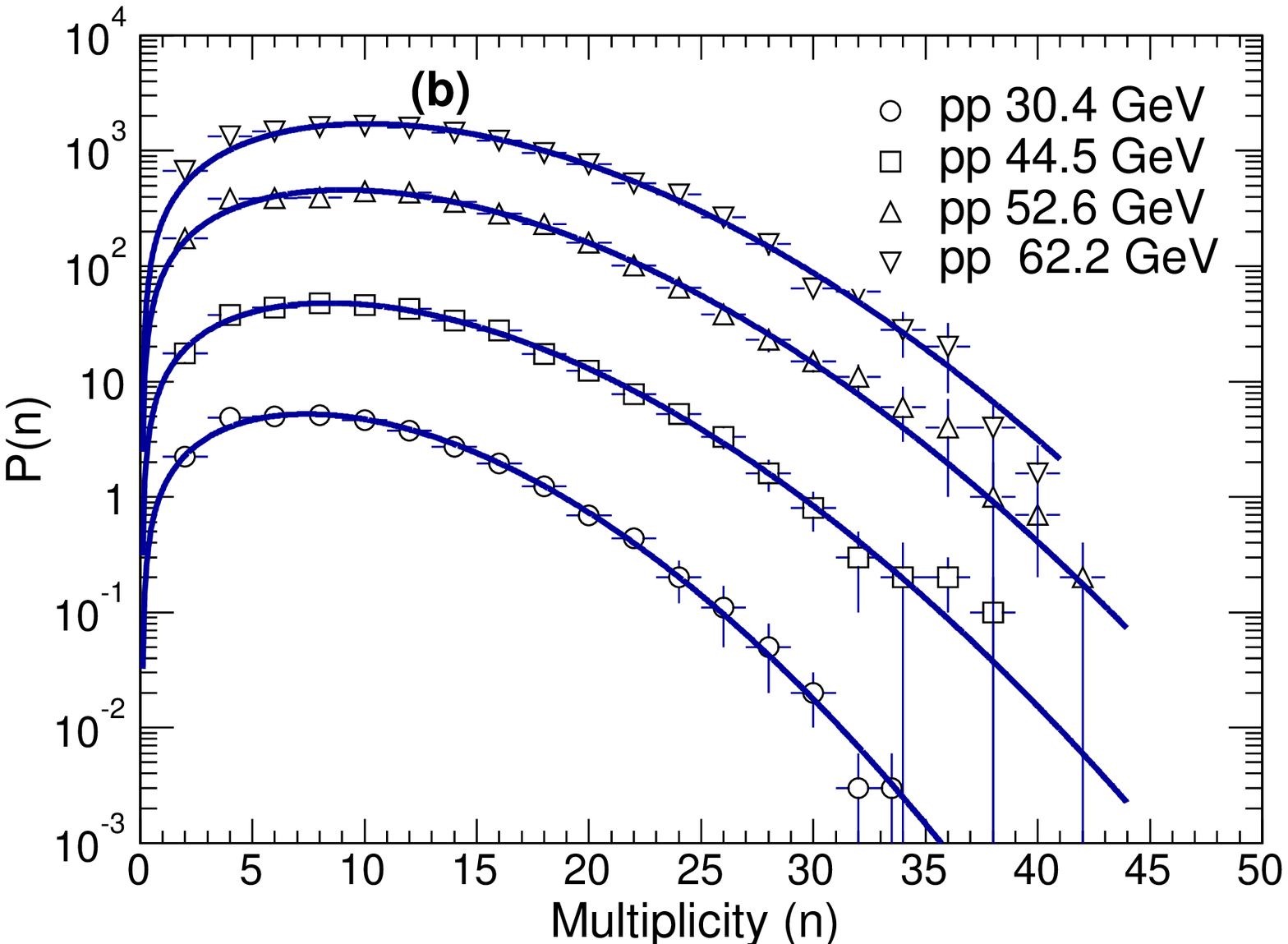}
\caption{ (a) LHCb \cite{18} measurement of charged particle multiplicity
  distribution in $p$-$p$ collisions at different pseudorapidity intervals at
  $\sqrt{s_{NN}}$ = 7 TeV (b) ISR measurement of topological
  cross-sections as a function of multiplicity in $p$-$p$ collisions at
  various energies \cite{19}.  The solid line represents the
  Weibull fit to the data points in (a) and (b). The data points for a given energy 
    are appropriately scaled for better visibility.}
\label{lhcball}
\end{figure}

The energy dependence of the mean charged multiplicity reflects the
underlying particle production mechanism. Feynman scaling predicts that the
total mean number of particles produced obeys an energy dependence
proportional to ln($\sqrt{s_{NN}}$) \cite{25}.
Figure 3(a) shows the  mean multiplicity as a function of the center of mass
energy.  As can be seen from the figure, the variation of  mean
multiplicity  with beam energy can be quantified by the expression \\

$\langle$ n $\rangle$  =  {\bf A} + {\bf B} [ln ($\sqrt{s_{NN}}$)] +
{\bf C} [ln$^{2}$ ( $\sqrt{s_{NN}}$ )] \\




We also observe that the scale parameter, $\lambda$ shows an energy 
dependence which is similar to that of the mean multiplicity . This is
shown in Figure 3(b).  We can see from Table 1 that the values of the
shape parameter $k$ do not vary significantly with the center of mass energy.
Taking this into account and the extrapolated values of the $\lambda$ 
parameter,  we predict the multiplicity  distribution at mid-rapidity
in $p$-$p$ collisions at the upcoming LHC run at $\sqrt{s_{NN}}$ = 5.5 TeV
($\lambda$ = 6.88, k = 1.15) and 14.0 TeV ($\lambda$ = 9.02 ,k = 1.15). 
Figure 4 depicts the predicted multiplicity distribution at
$\sqrt{s_{NN}}$ = 5.5 TeV and 14.0 TeV at mid-rapidity. The mean
multiplicity, $<n>$, at 5.5 TeV and 14.0 TeV  turns out to be 6.54 and
8.56, respectively.
The measurement of  multiplicity distribution at larger $\sqrt{s_{NN}}$ at
LHC  will be a first test of the further applicability of the Weibull
distribution and will add credibility to the extrapolation.

\begin{figure}
\includegraphics[scale=0.31]{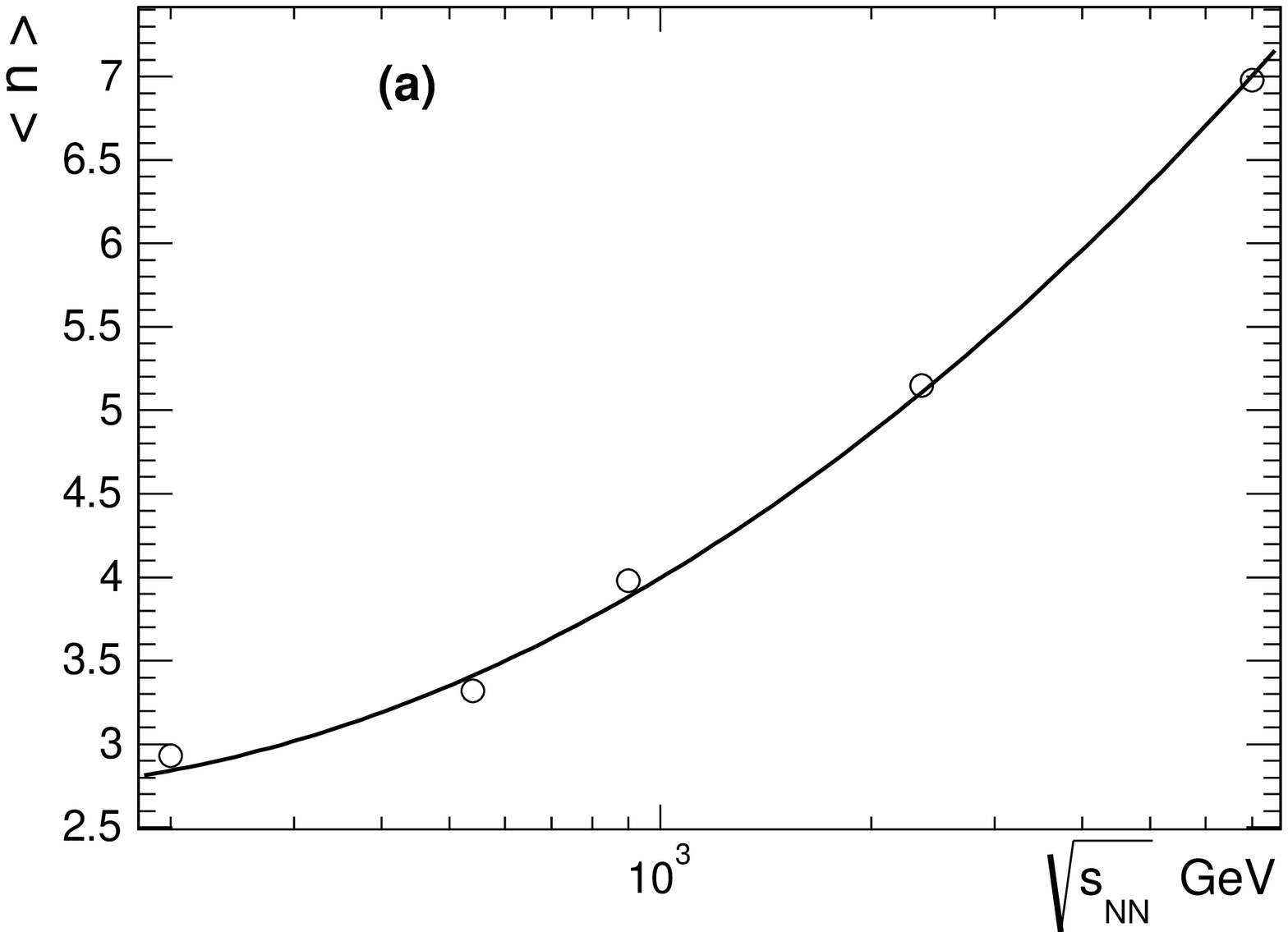}
\includegraphics[scale=0.31]{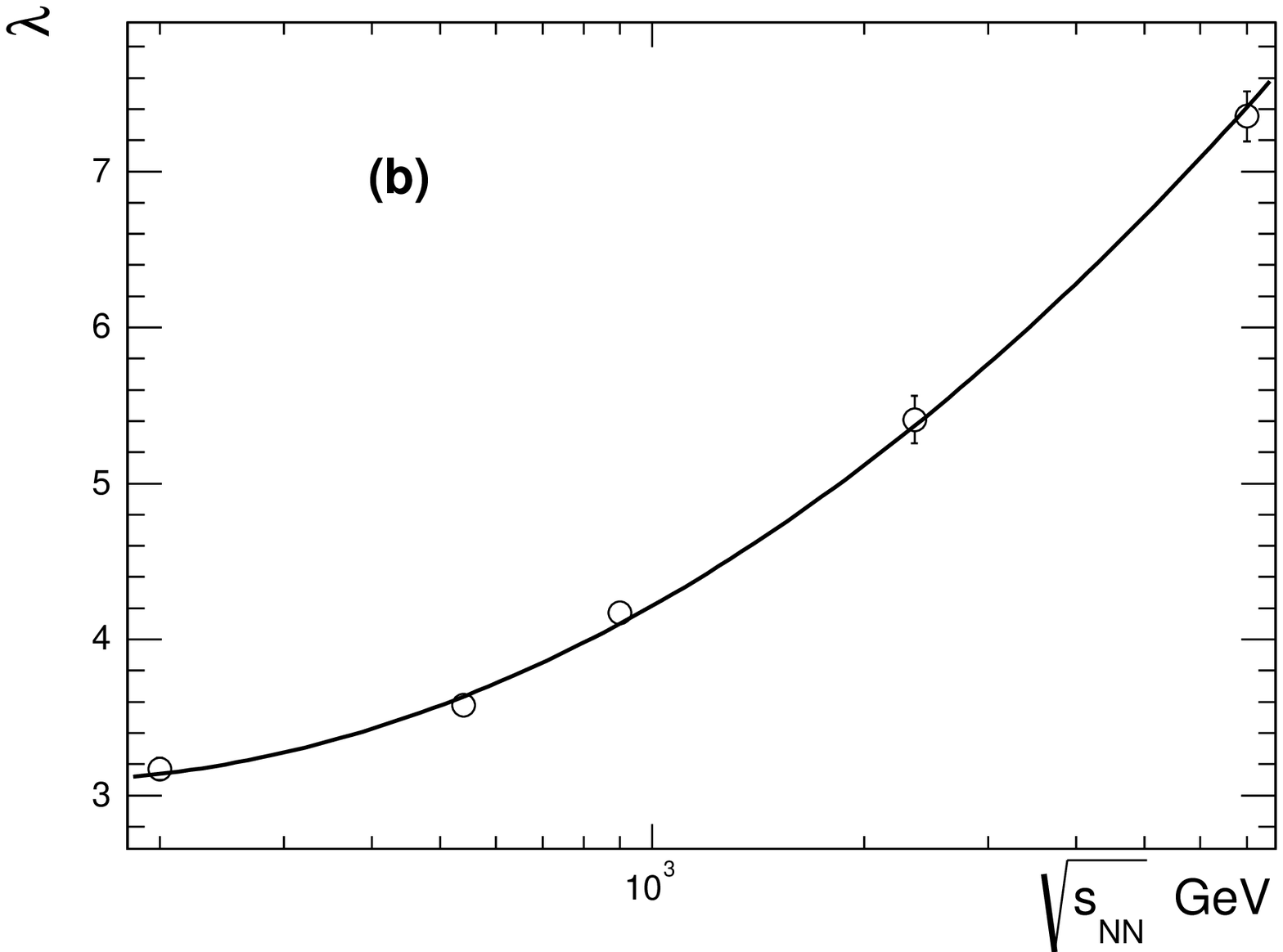}
\caption{ (a) The variation of mean multiplicity at mid-rapidity as a function of
  center of mass energy. (b) The variation of $\lambda$ parameter
  extracted from the Weibull fit as a function of beam energy  The
  solid line represents the fit given by expression (3) to the data
  points.}

\label{energy}
\end{figure}

\begin{figure}
\begin{center}
\includegraphics[scale=0.4]{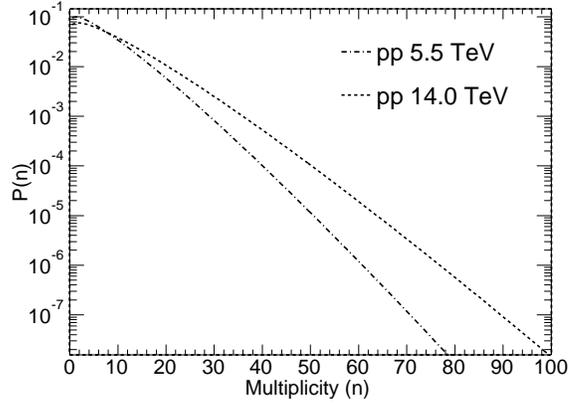}
\caption{The mid-rapidity charged particle multiplicity distribution
  in $p$-$p$ collisions at  $\sqrt{s_{NN}}$ = 5.5 TeV and 14.0 TeV as
  predicted by  Weibull parametrization.}
\label{meanpt}
\end{center}
\end{figure}
\section{Summary and Remarks}

We have demonstrated that the Weibull distribution provides an 
excellent description of the multiplicity distributions of inclusive charged
particles in hadronic collisions at all available energies and at all 
pseudorapidity intervals. This is particularly significant since the
Weibull distribution arises in cascade  processes involving
fragmentation of the source. This leads to a very interesting physics
interpretation in terms of current dynamical models of multi-particle
production. Because of the universal nature of the parton emission and
hadronization process, we claim that the Weibull distribution  gives the
most proper description compared to the previous distributions used
in multi-particle production processes.


\end{document}